\documentclass[a4paper,11pt]{article}
\usepackage{jheppub} % for details on the use of the package, please see the JINST-author-manual
\usepackage{comment}
\usepackage{lineno}
\title{\boldmath Unraveling xenon primary scintillation yield for cutting-edge rare event experiments}

\author[a,1]{C.A.O.~Henriques}

\author[a]{J.M.R.~Teixeira}

\author[a]{P.A.O.C.~Silva}

\author[a]{R.D.P.~Mano}

\author[a]{J.M.F. dos~Santos}

\author[a,1]{C.M.B.~Monteiro \note {Corresponding authors.}}

\affiliation[a]{LIBPhys, Physics Department, University of Coimbra, Rua Larga, Coimbra, 3004-516, Portugal}

\emailAdd{cristinam@uc.pt, henriques@uc.pt}

\abstract{
Xenon scintillation has been widely used in rare event detection
experiments, such as in neutrinoless double beta decay, double electron
captures and dark matter searches. Nonetheless, experimental values for primary scintillation yield in gaseous xenon (GXe) remain scarce and dispersed. The mean
energy required to produce a scintillation photon, \(w_{sc}\), in GXe in the absence of recombination has been measured to be in the range of
34--111 eV. Lower \(w_{sc}\)-values were reported for $\alpha$-particles
when compared to electrons produced by $\gamma$- or x-rays. Since \(w_{sc}\) is expected to be similar for x-, $\gamma$-rays or electrons and almost equal to that obtained for $\alpha$-particles, the above difference can not be understood. In addition, at present one may also pose the question of a dependence of \(w_{sc}\) on photon energy. 
We carried out a systematic study on the absolute primary scintillation yield in GXe under reduced electric fields in the 70--300 V cm\textsuperscript{-1} bar\textsuperscript{-1} range and near atmospheric pressure, 1.2 bar, supported by a robust geometrical efficiency simulation model.

We were able to clear-out the above standing problems: \(w_{sc}\) was determined for x/$\gamma$-rays in the 5.9–60 keV energy range as well as for $\alpha$-particles in the 1.5–2.5 MeV range, and no significant dependency neither on radiation type nor on energy has been observed. Our experimental \(w_{sc}\)-values agree well with both state-of-art simulations and literature data obtained for $\alpha$-particles. The discrepancy between our results and the experimental values found in the literature for x/$\gamma$-rays is discussed in this work and attributed to unaddressed large systematic errors in those previous studies. 
These findings can be extrapolated to other gases, and have impact on experiments such as double beta decay, double electron capture and directional dark matter searches while also on potential future detection systems such as DUNE-Gas.

Neglecting the 3\textsuperscript{rd} continuum emission, as is the case of most of the literature values, a mean \(w_{sc}\)-value of
38.7 \(\pm\) 0.6 (sta.)\(\ _{- 7.2}^{+ 7.7}\) (sys.) eV was obtained.
If the Xe 3\textsuperscript{rd} continuum
emission is to be considered, the average energy to produce a
2\textsuperscript{nd} and a 3\textsuperscript{rd} continuum photon was
calculated as \(w_{2^{nd}} =\) 43.5 \(\pm\) 0.7
(sta.)\(\ _{- 8.1}^{+ 8.7}\) (sys.) eV and \(w_{3^{rd}} =\) 483 \(\pm\)
7 (sta.)\(\ _{- 105}^{+ 110}\) (sys.) eV, respectively, while the energy
to produce either a 3\textsuperscript{rd} or a 2\textsuperscript{nd} continuum
photon is \(w_{2^{nd}{+ 3}^{rd}} =\) 39.9 \(\pm\) 0.6
(sta.)\(\ _{- 7.4}^{+ 8.0}\) (sys.) eV. 
 
}

\begin{document}
\maketitle
\flushbottom

\section{Introduction}
\label{sec:intro}

Gaseous xenon (GXe) is playing an increasingly significant role in
important areas of neutrino physics such as double beta decay and double
electron capture experiments \cite{NEXT:2019gtz,Obara:2019tnh,Gomez-Cadenas:2019ges,Galan:2019ake,Gavriljuk:2018pez,NEXT:2020sux}, and it may be used as detection medium in directional dark matter experiments as well as in MeV-region $\gamma$-ray imaging
\cite{nygren2013columnar,mimura2009xenon,10.1117/1.OE.53.2.021108}. The capability for simultaneous readout of both ionization
and scintillation signals and for topology reconstruction of the
ionizing particle tracks are important advantages of GXe. In addition,
GXe allows for improved energy resolution when compared to liquid xenon
(LXe) \cite{BOLOTNIKOV1997360} due to the observed fluctuations in energy deposition
between the ionization and the scintillation channels in LXe \cite{EXO-200:2003bso},
an effect that can be corrected to some extent by combining both
channels \cite{EXO-200:2012pdt}. Better energy resolution may lead to improved
electron/nuclear recoil discrimination, being also a major asset for the neutrinoless double beta decay sensitivity, and the topology of the
ionization track will be an additional tool, providing information about
the direction of the WIMP scatter or discrimination between single and double electron ionization tracks.

The precise knowledge of the xenon response to radiation interactions in
both scintillation and ionization channels is of utmost importance for
the exact understanding and modelling of the detector. The
primary scintillation yield of GXe is far less understood than the
ionization yield due to the limited number of studies existing in the literature.
Mimura \emph{et al.} \cite{10.1143/JJAP.48.076501} discusses the results obtained until then, while Serra \emph{et al.} \cite{10.1088/1748-0221/10/03/P03025} provides an update with further results published meanwhile.

The dominant scintillation mechanism in xenon at the atmospheric pressure is the so-called 2\textsuperscript{nd} continuum, a Gaussian-like emission spectrum, 10 nm in width. Its centroid has been reported in the 172--178 range \cite{10.1016/0167-5087(83)90028-5, 10.1016/0029-554X(79)90453-1, 10.1063/1.436349, 10.1063/1.440803, 10.1016/j.nima.2015.05.065, 10.1103/RevModPhys.82.2053}. In this work we consider the most recent value, 175 nm \cite{10.1016/j.nima.2015.05.065}. Nonetheless, other non-conventional scintillation mechanisms have been observed, such as the broadband Neutral Bremsstrahlung (NBrS) \cite{NBrS_prx} emission and the 3\textsuperscript{rd} continuum emission in the 250–400 nm range \cite{10.1063/1.436349}. All the \(w_{sc}\)-values present in the literature, except for the most recent study carried out for $\alpha$-particles \cite{10.1140/epjc/s10052-022-10385-y}, consider the 3\textsuperscript{rd} continuum and neutral bremsstrahlung emissions to be negligible. 

At reduced electric
fields, i.e. the electric field normalized by the gas pressure,
$E/p$, above 60 V cm\textsuperscript{-1}
bar\textsuperscript{-1}, the recombination of primary electrons/ions
produced during the radiation interaction is negligible \cite{10.1088/1748-0221/10/03/P03025, 10.1088/1748-0221/8/05/P05025, 10.1143/JJAP.48.076501, 10.1109/tns.2004.836030}. In those conditions, the average energy required to excite a xenon atom,
\(w_{ex}\), is similar to the average energy expended per
scintillation photon $w_{sc} = E_{dep}/N_{ph}$, where
\(N_{ph}\) is the number of scintillation photons and $E_{dep}$ the deposited energy \cite{10.1088/1748-0221/10/03/P03025,10.1088/1748-0221/8/05/P05025}.
\(w_{ex}\) does not depend on gas density, below 0.2
g/cm\textsuperscript{3}, $\sim$20 bar at room temperature \cite{10.1143/JJAP.48.076501}.

Several measurements of the \(w_{sc}\) for x- and $\gamma$-ray interactions have become available, although with highly dispersed values from 61 to 111 eV \cite{10.1088/1748-0221/5/09/P09006, 10.1088/1748-0221/3/07/P07004, 10.1016/j.nima.2023.168038, 10.1109/23.106674, 10.1016/j.nima.2015.04.057}. The \(w_{sc}\)-values obtained for 5.5-MeV $\alpha$-particle interactions are less dispersed, in the 34--60 eV range, and $\sim40$\% lower on average than its x/$\gamma$-ray counterpart despite being obtained for similar working conditions \cite{10.1088/1748-0221/10/03/P03025, 10.1143/JJAP.48.076501, 10.1109/TNS.2003.820615, 10.1109/tns.2004.836030, 10.1140/epjc/s10052-022-10385-y}. An overview of the \(w_{sc}\)-values present in the literature can be found in Table \ref{tab:literature}.

The average energy expended per excited atom in GXe is expected to be similar for x-, $\gamma$-rays or electrons and almost equal to that obtained for $\alpha$-particles \cite{10.1143/JJAP.48.076501}. However, results presented in the literature are inconsistent with that expectation. The difference between the above results is presently not fully understood, as can be only partially ascribed to the different gas density and/or drift field conditions. In addition, at present one may also pose the question of a dependence of \(w_{sc}\) on photon energy.

These inconsistencies motivated us to pursue further experimental studies in a dedicated setup where the primary scintillation could be isolated and studied in detail. In this work, we report new results on the xenon \(w_{sc}\)-value in absence of recombination, for $\alpha$-particles in the 1.5- to 2.5-MeV range and for x/$\gamma$-rays in the 6--60-keV range. These findings can be extrapolated to other noble gases, like Ar and Kr, and might have also impact on potential future detectors such as DUNE-Gas \cite{10.48550/arXiv.2203.06281}.

\begin{table*}[tbp]
\setlength{\tabcolsep}{11pt}
\renewcommand{\arraystretch}{1.3}
\centering
\begin{tabular}{c c c c}
\hline \hline
\bf \boldmath $w_{sc} $ (eV) & \bf radiation & \bf energy (keV) &  \bf observations \\
\hline 
74 \(\pm\) 10 & x-rays & 5.9 & 1 bar, \cite{10.1088/1748-0221/5/09/P09006} \\
111 \(\pm\) 16 & x-rays & 5.9 & 1 bar, \cite{10.1088/1748-0221/3/07/P07004} \\
80 \(\pm\) 12 & x-rays & 5.9 & 1 bar, \cite{10.1016/j.nima.2023.168038} \\
76 \(\pm\) 12 & $\gamma$-rays & 59.5 & 15 bar, \cite{10.1109/23.106674} \\
61 \(\pm\) 18 & $\gamma$-rays & 662 & 14 bar, \cite{10.1016/j.nima.2015.04.057} \\
39.2 \(\pm\) 3.2 & $\alpha$-particles & 5486 & 10 bar \cite{10.1088/1748-0221/10/03/P03025} \\
34.1 \(\pm\) 2.4 & $\alpha$-particles & 5486 & 5 bar \cite{10.1143/JJAP.48.076501} \\
34.5 \(\pm\) 1.7 & $\alpha$-particles & 5486 & 1-10 bar \cite{10.1109/TNS.2003.820615} \\
59.4 \(\pm\) 1.5 & $\alpha$-particles & 5486 & 1-2 bar \cite{10.1109/tns.2004.836030} \\
50.5 \(\pm\) 5.9 & $\alpha$-particles & 5486 & 1 bar, VUV \cite{10.1140/epjc/s10052-022-10385-y} \\
\hline
\hline
\end{tabular}
\caption{\label{tab:literature}A summary of \(w_{sc}\)-values presented in the literature. All values are
considered to be obtained with negligible electron-ion recombination,
since authors used reduced electric fields of at least 70 V
cm\textsuperscript{-1} bar\textsuperscript{-1}. Some relevant working conditions such as the gas pressure, radiation type and energy are also listed.}
\end{table*}

\section{Experimental setup}
\label{sec:setup}

\begin{figure}[tbp]
\centering % \begin{center}/\end{center} takes some additional vertical space
\includegraphics[width=14cm]{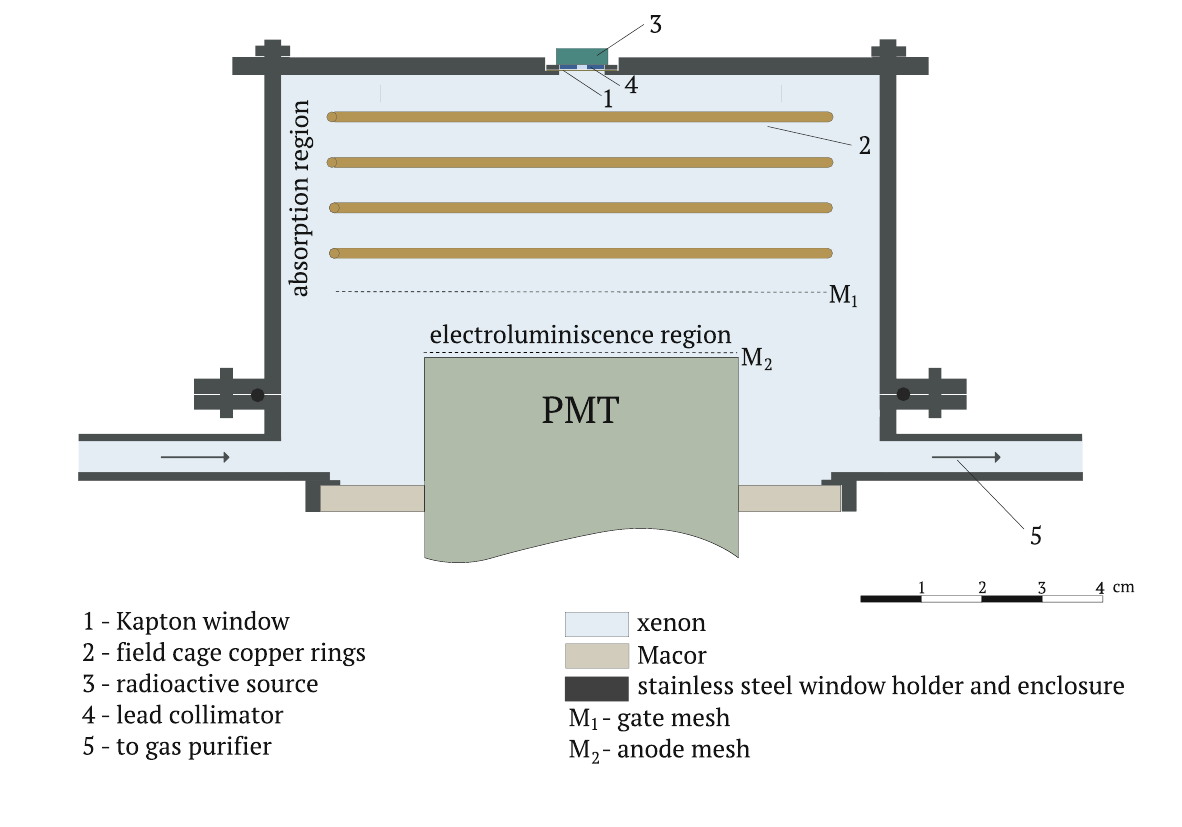}
\caption{\label{fig:setup} Schematic of the gas scintillation chamber used in this work.}
\end{figure}

For detailed studies of the primary scintillation, we built a gas
proportional scintillation chamber (GPSC) with photomultiplier tube (PMT)
readout, Fig. \ref{fig:setup}.
This apparatus allows us to study both primary and secondary
scintillation signals, S1 and S2, respectively. The primary
scintillation occurs mostly in the absorption region, where a weak
electric field prevents electron-ion recombination and guides the
ionization electrons toward a stronger electric field region, where the
secondary scintillation produced by electron impact, also known as electroluminescence (EL), takes
place.

The absorption region is
delimited by a Kapton window, 12-$\mu$m thick, aluminized
on the inner side, and the gate wire mesh, 0.9-mm pitch stainless-steel
wires with 80 $\mu$m in diameter. The anode is made from a similar wire
mesh, placed above the photosensor. The chamber, filled with
$\sim$1.2 bar of xenon, consists of a stainless-steel cylinder, 10 cm in
diameter, the bottom part of the body being a Macor disc epoxied
to the PMT and to the detector wall for electrical insulation.
The GPSC is operated at room temperature, with the xenon gas circulating
by convection through St707 SAES getters \cite{saes}, heated to
temperatures up to 250 \({^\circ}\)C. A field cage of four equally-spaced copper rings interconnected through resistors was assembled in the absorption
region to ensure electric field uniformity. A polytetrafluoroethylene (PTFE) structure supports
the copper rings. Electric field maps of the GPSC were obtained using a finite element method solver \cite{elmer}. The electric field was found to vary by 8\% along the 3.6-cm thick absorption region and by 0.7\% along the 1-cm electroluminescence region. These values are conservative, as they represent the maximum field variation in a 7-mm radius cylindrical volume, where 95\% of the transversely diffused electrons are contained within.

\begin{figure}[tbp]
\centering % \begin{center}/\end{center} takes some additional vertical space
\includegraphics[width=7.5cm]{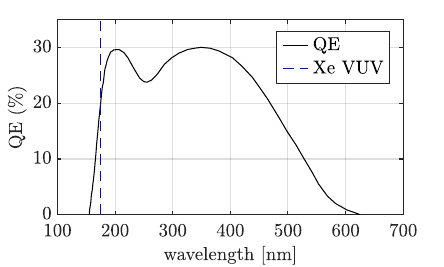}
\caption{\label{fig:QE} Quantum efficiency (QE) curve of the PMT used in this
work. The Xe VUV scintillation wavelength is also represented, assumed to be 175 nm \cite{10.1016/j.nima.2015.05.065}.}
\end{figure}

The PMT is an eight dynode model EMI D676QB with a diameter of 52 mm,
an effective cathode diameter of 45 mm, and a spectral sensitivity in
the range of 155--625 nm. The PMT quantum efficiency (QE) curve as
provided by the manufacturer is depicted in Fig. \ref{fig:QE}. Accordingly, a QE of $(20.9 \pm 1.5)\%$,  is expected for the Xe 2\textsuperscript{nd} continuum, the error stemming from
the differences between emission spectra in the literature.

The PMT signals are directly recorded with a WaveRunner 610Zi oscilloscope
from LeCroy, featuring a sampling rate of 10 GS/s. The PMT output is
connected to a load resistor of 200-Ω to convert the signal current into
voltage. A 50-Ω resistor would be preferable to match the cable
impedance and to reduce wave reflections. However, due to the low gain of
our PMT, $\sim 10^{5}$, a higher resistor value was required to increase the
signal-to-noise ratio up to acceptable levels. Nonetheless, wave
reflections in the cable terminations are properly handled, as will be
explained in the following sections.

Four radioactive sources, \textsuperscript{109}Cd,
\textsuperscript{244}Cm, \textsuperscript{241}Am, and
\textsuperscript{55}Fe having the 6.4 keV x-rays removed by means of a
chromium filter, collimated up to 2 mm and positioned 1 mm
above the detector window, were used in the present study to produce
x/$\gamma$-rays in the range of 5.9--60 keV and 5.5-MeV alpha particles (being
\(\sim\)2 MeV deposited in the Xe gas). In addition, a Tb target was
irradiated with 59.5 $\gamma$-rays to provide fluorescence x-rays in the 14-50 keV range. Overall, the counting rate was kept around 10 Hz to avoid pulse pileup, maintaining the cleanliness of the waveform baseline.

\section{Monte Carlo simulations}
\label{sec:simulation}

Despite the large solid angle subtended by the PMT with respect to the
secondary scintillation region, most of the primary scintillation
photons are produced near the detector window, far away from the
photosensor (see Fig. \ref{fig:setup}). Consequently, photon reflections on the
detector materials play an important role in the optical geometrical
efficiency. Since this parameter is crucial to estimate the number of
emitted primary and secondary scintillation photons, we developed a
detailed optical simulation of the detector geometrical efficiency (GE)
using the \textsc{GEANT4} toolkit \cite{10.1016/j.nima.2016.06.125}.

Detector components were designed and meshed with the 
software \textsc{FreeCAD}. Components made from the same materials were grouped
and their CAD geometries were directly imported into \textsc{GEANT4} using the
open-source \textsc{CADMesh} header \cite{10.1007/s13246-012-0159-8}.
Optical processes such as photon reflection and refraction are handled
by the \textsc{G4OpticalPhysics} class. The \textsc{GEANT4} simulation was
developed to account for the full wavelength range from 150 nm to 650
nm, allowing to study other light emission mechanisms beyond the Xe
2\textsuperscript{nd} second continuum, such as the
3\textsuperscript{rd} continuum and the neutral bremsstrahlung \cite{10.1140/epjc/s10052-022-10385-y, NBrS_prx}. The refractive indexes of the fused
silica PMT window and Xe gas were implemented as a function of
photon wavelength, allowing for full simulation of Fresnel reflections
and refractions \cite{10.1364/JOSA.55.001205,
10.1109/tdei.2006.1657967}. The \textsc{glisur} model was used for
boundary processes, and the respective optical surface properties, e.g.
reflectivity, absorption, and finishing, were defined. Some of the most
relevant parameters comprise the reflectivity of both aluminium and
stainless steel; the former covers the detector window inner surface,
and the latter is used in most of the inner surfaces, including the gate
and anode meshes. Experimental reflectivity values measured in
\cite{10.1016/j.nuclphysbps.2007.08.059} were considered for the
2\textsuperscript{nd} continuum of xenon scintillation, while the remaining
wavelength region was covered with values reported in
\cite{10.1364/JOSA.51.000719, 10.1016/0165-1633(89)90039-7}. Less relevant materials like
the PMT Macor holder, the field cage's copper rings and the PTFE
structure were also included in the \textsc{GEANT4} simulation, their
reflectivity values being taken from \cite{10.1063/1.3318681, silva2010reflection, macor}.

\begin{figure}[tbp]
\centering\centering\centering
\includegraphics[width=12cm]{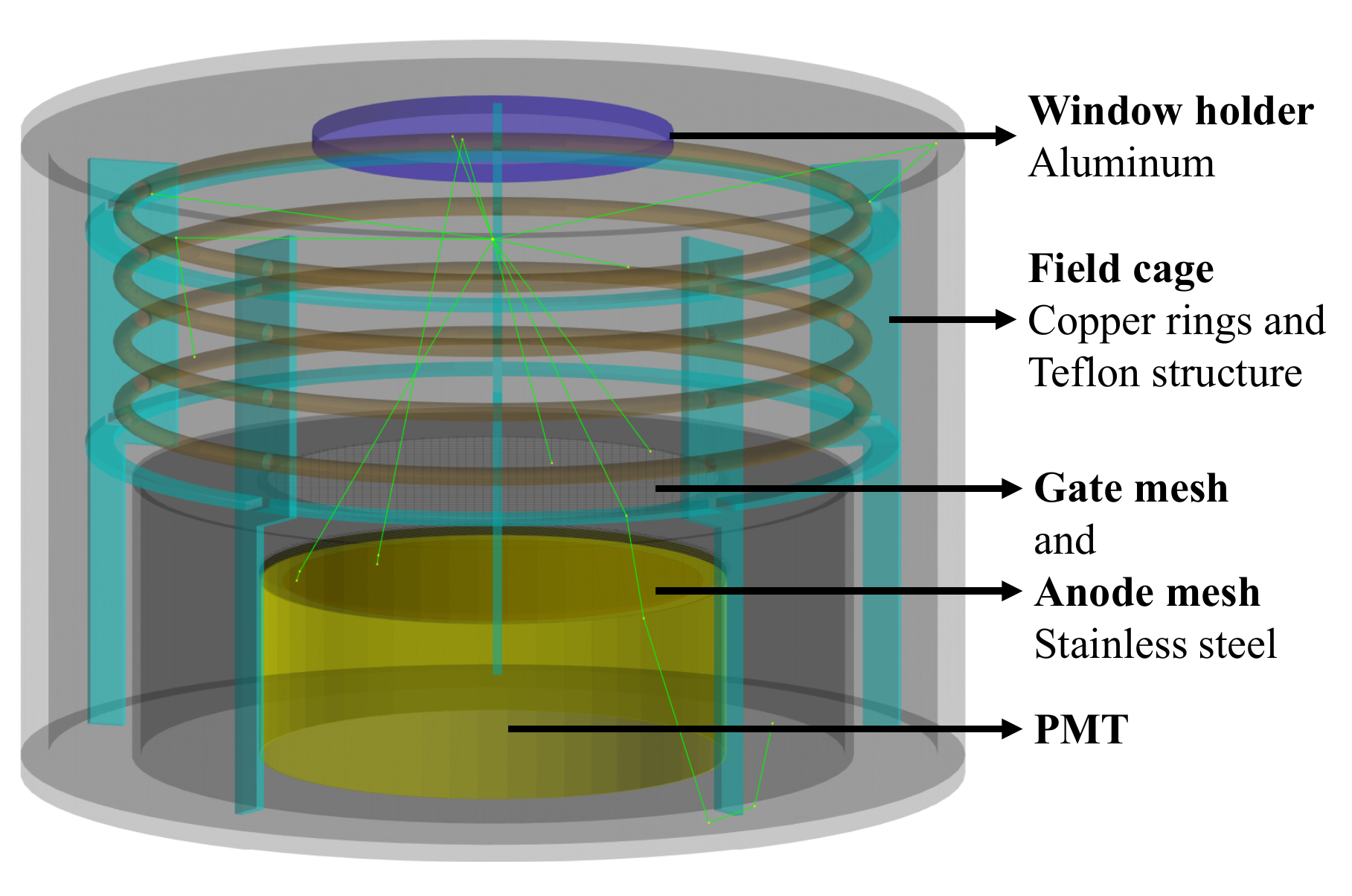}
\caption{\label{fig:geant4}A 3-dimentional representation of the \textsc{GEANT4} optical
simulation, showing the raytracing of 10 photons.}
\end{figure}

Figure~\ref{fig:geant4} shows a 3-dimensional view of the \textsc{GEANT4}
simulation model and the ray-tracing of 10 scintillation photons
generated 36 mm away from the PMT. The detector GE along the central axis perpendicular to the
PMT can be found in Figure~\ref{fig:GE}. The GE refers to the ratio of photons reaching the PMT photocathode, i.e. a sensitive area placed below the PMT window, when photons are
generated with random directions at different distances from the PMT.
The impact of radial effects, such as
the incident beam divergence, the size of the ionization electron cluster
and the electron transverse diffusion of the ionization electron cloud have been taken into account.

\subsection{Primary scintillation emission}
\label{sec:simS1}

\begin{figure}[tbp]
\centering\centering\centering
\includegraphics[width=7.5cm]{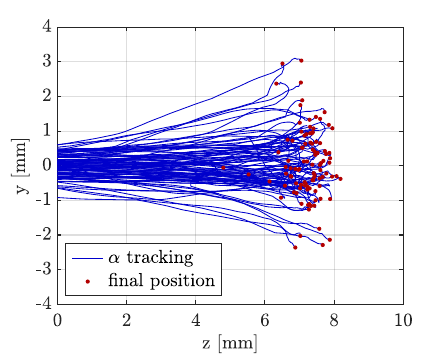}
\includegraphics[width=7.5cm]{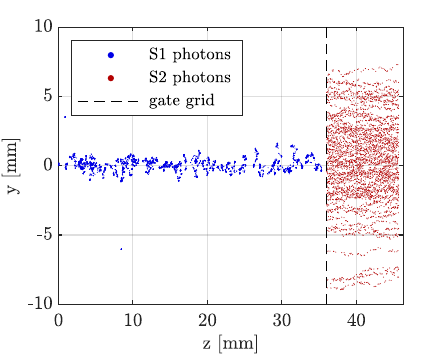}
\caption{\label{fig:tracking}(Left) Tracking of 100 $\alpha$-particles in xenon gas
at 1.2 bar, as obtained from \textsc{GEANT4}; (right) the position distribution
of S1 and S2 photons produced by 22.1-keV x-ray events in xenon at 1.2
bar and $E/p$ values of 0.15 and 2.3 kV cm\textsuperscript{-1} bar\textsuperscript{-1} in the absorption and EL region, respectively, as obtained from the combination of \textsc{GEANT4}, \textsc{Degrad} and \textsc{Garfield++}. The S1 distribution comprises 100 x-ray interactions, while the S2 data stem from 100 electrons randomly selected from a much larger sample. For clarity, only 10\% and 20\% of S1 and S2 photons, respectively, are represented.}
\end{figure}

The beam divergence of $\alpha$-particles and x/$\gamma$-rays was obtained from a
non-optical \textsc{GEANT4} simulation, considering the aperture of the
collimators used in the experimental campaign. Alpha-particles and
x/$\gamma$-rays are generated isotropically at random positions above the
collimator and tracked. For $\alpha$-particles the energy deposition along the
track is computed with a small step size and used to estimate the
distribution of primary photons and ionization electrons produced in the xenon gas.
Figure~\ref{fig:tracking} left shows an example of $\alpha$-particle
tracks inside the gas volume. However, for x/$\gamma$-rays we rely on the more
accurate \textsc{Degrad} model, a \textsc{Fortran} toolkit developed by S. Biagi \cite{degrad}, to simulate the primary cluster of xenon excited states and ionization
electrons, including the shell absorption by photoelectric effect and subsequent Auger, shake-off and fluorescence emission. For this
reason, the \textsc{GEANT4} tracking of x/$\gamma$-rays is stopped as soon as they
interact inside the xenon volume, the interaction positions being
recorded. This is only possible for x/$\gamma$-rays because $\alpha$-particles are not
supported by \textsc{Degrad}.

The working conditions used in our experimental campaign were assumed in
\textsc{Degrad}, including the electric field values, pure xenon pressure of 1.2 bar, and
estimated temperature of 296 K. Incident x-rays with energies in the
5.9-59.5 keV range were simulated and the total number of primary
electrons and xenon excitations per event were obtained, along with their
(x,y,z) coordinates. The overall distributions of the primary
scintillation photons and electrons are computed, combining the
distributions of xenon excited states and electrons obtained from \textsc{Degrad}
with the x/$\gamma$-ray interaction positions obtained from \textsc{GEANT4}. Figure
\ref{fig:tracking} right shows an example of the positions where primary scintillation photons are generated for 22.1-keV x-rays.

Finally, the \textsc{GEANT4} optical simulation is supplied with a randomized sample
of photons emitted isotropically following the aforementioned
distributions, allowing to compute the GE curve in the absorption
region.

\subsection{Secondary scintillation emission}
\label{sec:simS2}

The cluster of ionization electrons obtained for $\alpha$-particles from \textsc{GEANT4}
and for x/$\gamma$-rays from the \textsc{GEANT4-Degrad} combination are 
imported into \textsc{Garfield++}, a toolkit for the detailed simulation 
of detectors that use gases or semi-conductors as the sensitive media
\cite{garfieldpp}.
\textsc{Garfield++} provides an interface to \textsc{Magboltz}, a simulation tool that computes the electron transport parameters in the gas \cite{Magboltz}. In \textsc{Garfield++} the geometry 
is defined using the class \textsc{GeometrySimple} and the uniform electric field
is set using the \textsc{ComponentConstant}. This is done independently for the
absorption and the EL region, allowing to set different drift models for
computational time requirements, enabling to achieve a better balance between computational efficiency and accuracy of the simulation results. Gas parameters are also set for both
regions using the class \textsc{MediumMagboltz}. The drift path is
computed using Monte Carlo integration with the class
\textsc{AvalancheMC}. After the electrons drift throughout the absorption region, their final positions are imported to the electroluminescence
region where the electric field value is higher than the gas scintillation threshold. The movement of the electrons in this
region is modelled using the class \textsc{AvalancheMicroscopic}, where
the electron is tracked from collision to collision. This model enables
the detailed calculation of ionisation and excitation processes. With
this method both the coordinates and the number of excited xenon atoms are
retrieved and subsequently imported into \textsc{GEANT4} where the optical simulation
takes place. Accordingly, the radial distribution of the excited states of xenon in
the EL region obtained from \textsc{Garfield++} accounts for the x/$\gamma$-ray beam
divergence, the initial electron cloud size and the transverse electron
diffusion in both regions. Figure~\ref{fig:tracking} also shows an
example of a distribution for xenon excited states in the EL region simulated
for 22.1-keV x-rays.

\subsection{Geometrical efficiency simulation results}
\label{sec:GE}

\begin{figure}[tbp]
\centering\centering\centering
\includegraphics[width=10cm]{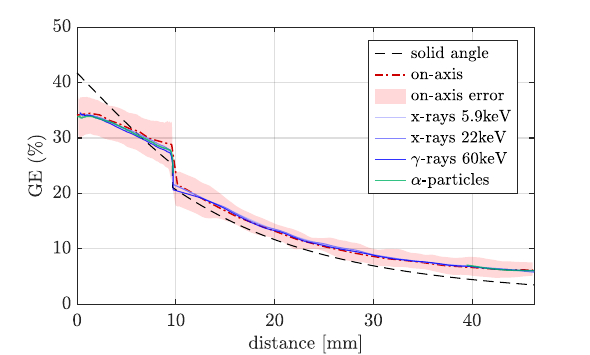}
\caption{\label{fig:GE}GE simulation results for photon emission as a 
function of distance to the PMT window, considering the on-axis photon emission and the 3-dimensional photon
emission when taking into account radial effects, such as beam
divergence, electron cluster size and transverse electron
diffusion. The GE curve calculated considering only
the solid angle and the transparency of gate and
anode meshes is also depicted for comparison.
The anode and gate meshes were placed at 0 and 10 mm, respectively.}
\end{figure}

Figure~\ref{fig:GE} depicts the \textsc{GEANT4} GE simulation results for photon emission as a function of distance to the PMT window. For obtaining the GE curves two different cases were considered, namely assuming the on-axis approximation and, for $\alpha$-particles and for three different x/$\gamma$-ray
energies, considering radial effects, taking into account the respective collimator sizes and experimental working conditions. The simplistic GE curve calculated considering only the solid angle and the transparency of gate and
anode meshes is also depicted for comparison.

As expected, calculations based solely on solid angle considerations are
highly inaccurate. The GE near the
detector window can be underestimated by almost 50\% due to the relatively higher contribution of photon
reflections. Moreover, the simplistic assumption of the mesh transparency fails when neglecting photons hitting the mesh
planes at wide angles. Figure~\ref{fig:GE} also shows the 68\%-confidence level error of
the \textsc{GEANT4} on-axis simulation, which is dominated by the uncertainty in reflectivity and unknown polishing degree of some materials.

The degradation of the GE arising from radial effects is about 5\% in
the EL region and negligible in the absorption region. 
Remarkably, the GE curves obtained for different x/$\gamma$-ray energies and
$\alpha$-particles are comparable even in the EL region. For simplicity, the four GE curves 
considering radial effects depicted in Fig.~\ref{fig:GE} are averaged, the result being used in further GE corrections.

\section{Analysis methodology}
\label{sec:analysis}

\subsection{Waveforms sampling and pre-processing}
\label{sec:sampling}

The PMT signal is split into two oscilloscope channels: a
``full-channel'' that is used to record the S2 pulse without saturation,
and a ``zoomed-channel'' that is optimized for S1 measurements. The signal of the oscilloscope is triggered on the ``full-channel'' using the rising edge of the S2 pulse. This
setup allows us to measure simultaneously the S2 and S1 pulses with high
amplitude resolution. Moreover, the ``zoomed-channel'' delivers lower
baseline fluctuations dominated by the oscilloscope electronic
noise.

PMT waveforms have been sampled at 5--10 GS/s. Nevertheless, we were
forced to compress the data due to memory limitations in larger x/$\gamma$-ray
acquisitions. This was accomplished by averaging successive waveform
data points at periodic intervals. On the one hand, such a procedure
degrades the time resolution of waveforms, from 0.1--0.2 ns to 10--70 ns,
which is still adequate for our studies. On the other hand, it increases
the amplitude resolution beyond the 8-bit limit of the
oscilloscope. For convenience, the PMT waveforms are inverted, i.e.,
waveforms presented from now on have positive amplitude, despite the PMT
signal being negative.

\begin{figure}[tbp]
\centering % \begin{center}/\end{center} takes some additional vertical space
\includegraphics[width=12cm]{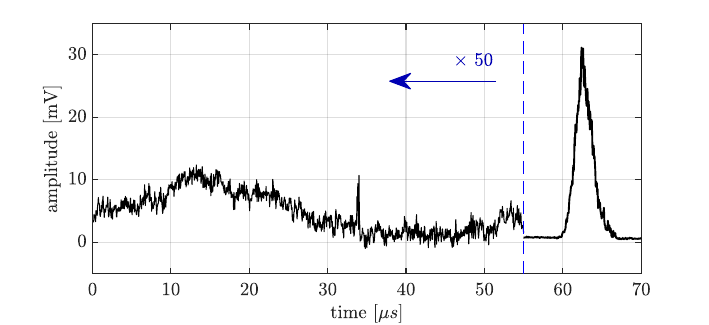}
\caption{\label{fig:single_wf} Typical PMT waveform obtained for 5.9-keV x-rays,
showing both the primary and the secondary signals, S1 and S2.}
\end{figure}

Figure~\ref{fig:single_wf} depicts a typical, already compressed PMT
waveform obtained for 5.9-keV x-rays. The large pulse around 63 $\mu$s is
the S2 pulse of an x-ray event, and the small spike around 34 $\mu$s is
likely an S1 single photoelectron. The x-ray and gamma-ray energy range, from 6 to 60 keV,
studied in this work produce on average only 1 to 15
photoelectrons per interaction, which combined with the low gain of
our PMT makes the primary scintillation yield difficult to quantify in a per-event basis.
Therefore, despite the S1 footprint being clearly seen in the example of
Fig.~\ref{fig:single_wf}, it might be indistinguishable from the
electronic noise for other events. For this reason, we rely on waveform
averaging to cancel out the electronic noise hence revealing the
primary scintillation signal. 
Low-frequency fluctuations of the
oscilloscope baseline are the major source of statistical error in the S1
measurements. Yet, this can be mitigated with a sufficiently
large accumulation of waveforms. Each of the data runs comprises between 
10\textsuperscript{5}  and 2$\times$10\textsuperscript{6} 
waveforms, and can span 1 to 3 days of continuous data
acquisition. The stability of the system has been continuously monitored
during such long acquisition periods, by surveilling the centroid of the highest peak in
the energy spectrum of the radioactive source being in use, obtained
from the S2 pulse integration.

A pre-processing algorithm was developed to discriminate background
events (e.g., cosmic radiation), as well as waveforms with features that are unsuitable for
further analysis. This is particularly important for the accuracy of
primary scintillation measurements. Since S2 pulses are 4 orders of
magnitude larger than S1 pulses, a single background event can
jeopardize the entire sample of events over which S1 is averaged.
An example is a waveform with two S2 pulses for which one of them falls into
the S1 region. Therefore, the discrimination algorithm rejects waveforms
based on the baseline cleanliness, as well as on the duration, time offset and
shape of the S2 pulse. In addition, the oscilloscope baseline offset is
measured at the beginning of every recorded waveform and corrected
accordingly.

\subsection{PMT calibration}
\label{sec:PMTgain}

An accurate calibration of the photosensor gain is crucial to quantify
both primary and secondary scintillation yields. A blue LED biased with
direct current was used to obtain the single photoelectron charge
distribution of the PMT. Since the detector radiation window can
transmit visible light into the PMT, there was no need to place the
LED inside the gas chamber. 
PMT waveforms were digitized with a sampling
rate of 10 GS/s, allowing to resolve the short pulses of
single-photoelectrons. The oscilloscope trigger threshold was set as low
as possible without being overly saturated by background spikes. The LED
light intensity was adjusted for the probability of having more than
one photoelectron event in the same waveform to be negligible.
Several measurements with different trigger levels and LED intensities
were carried out to access the experimental uncertainty. The PMT gain
calibration was performed several times along the full experimental
campaign to monitor its performance. The gain variation was found to be within experimental errors.

\begin{figure}[tbp]
\centering
\includegraphics[width=7.5cm]{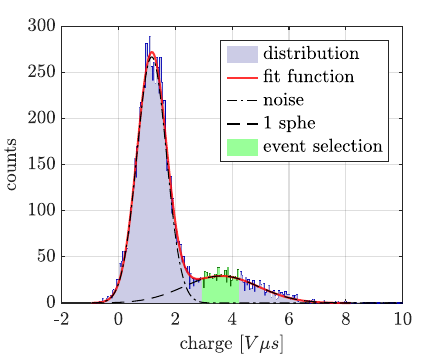}
\includegraphics[width=7.5cm]{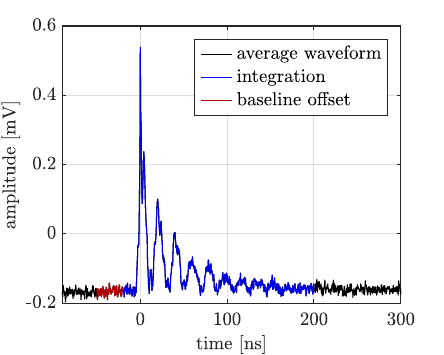}
\caption{\label{fig:gain2} (Left) Charge distribution of the first pulse
of single-photoelectron waveforms integrated between -2 and 7 ns. The corresponding fit function is also depicted, along with the selected region of single-photoelectron events. Only the electronic noise and single-photoelectron Gaussian functions are represented, as the double-photoelectron contribution is not visually perceptible; (right) average waveform for the
events selected in the left panel. The integration regions of the
baseline offset and the photoelectron signal are also shown.}
\end{figure}

Since a 200-Ω load resistor is used to collect the anode signal, wave reflections at both ends of the signal cable are significant and needed to be account for. However, due to the relatively large baseline fluctuations, waveform integration of the single photoelectron pulse for time durations above 200 ns, necessary to take into account those reflections, results in a large noise peak in the PMT charge distribution, engulfing the single photoelectron peak. Therefore, instead of integrating the full single photoelectron signal, only a short region of the waveform containing the first pulse was integrated, between -2 and 7 ns with respect to the oscilloscope trigger time. This small integration time alleviates the impact of the baseline fluctuations, with the disadvantage of excluding the reflected pulses. 
Figure~\ref{fig:gain2} left shows the charge distribution resulting from that integration. A sum of three Gaussians was fitted to this charge distribution: the first Gaussian
accounts for the electronic noise with area, centroid, and sigma being
left as free parameters, while the other two Gaussians account for
single- and double-photoelectron emission. Their centroids follow the
scaling \(1pe\) and \(2pe\), respectively, where \(pe\) is the centroid
of the single-photoelectron Gaussian, with standard deviations
\(\sigma\) following \(\sigma\sqrt{1}\) and \(\sigma\sqrt{2}\),
respectively, and the areas being related through Poisson statistics.
The centroid and the standard deviation of the single-photoelectron
Gaussian were left as free parameters, as well as the rate parameter, $\lambda$, of
the Poisson distribution.

This charge distribution was, then, used to select a sample of single-photoelectron events, with charge values between \(pe - 0.5\sigma\) and \(pe + 0.5\sigma\), falling outside the electronic noise and the double-photoelectron Gaussians, Fig.~\ref{fig:gain2} left. The waveforms of these selected events are then averaged and the baseline offset, measured before the photoelectron pulse, subtracted. 
Figure \ref{fig:gain2} right shows a typical average waveform obtained from those selected events, highlighting the full pulse and the baseline offset regions. The chain of reflected pulses is, now, clear, due to the cancellation of the electronic noise. As expected, the spacing between pulses, \(\sim\)16 ns, corresponds to twice the cable delay, which is 8 ns. The full photoelectron signal was integrated between -10 and 200 ns. A mean charge value per single photoelectron of 10.9 \(\pm\) 0.3 mV ns was obtained for a PMT biasing of 1450 V.

As a crosscheck for the above value, a second method was used: subtracting previously the baseline to each individual waveform, by considering its baseline as a straight line defined by the two offset values in the regions just before the starting of the pulse and just after the 200-ns pulse duration, and integrating the full waveform. Averaging over all waveform integrals, one obtains a charge value 4\% lower than what was obtained by the former method, being this difference  most likely due to the baseline correction constraints.

We note that the probability of double-photoelectron emission from the PMT
photocathode is negligible in the visible region, but may reach 20\%
for VUV photons \cite{Faham:2015kqa}. Accordingly, one could expect that our
experimental results would also be affected by the double-photoelectron effect.
However, this contribution is cancelled out in calculations, since we use the PMT QE curve provided by the manufacturer, which also includes
this effect.

\begin{figure}[tbp]
\centering
\includegraphics[width=12cm]{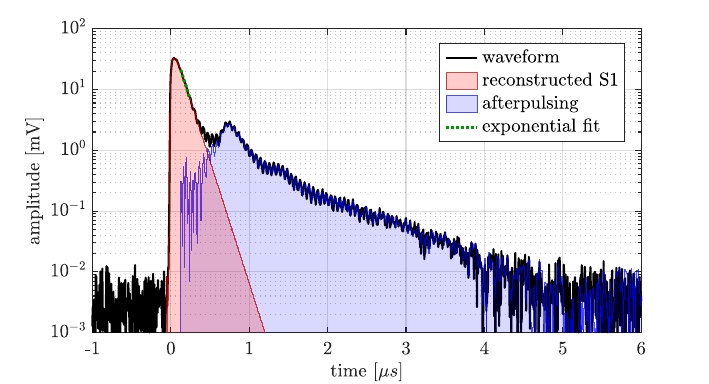}
\caption{\label{fig:afterpulsing} Average wavefrom obtained from $\alpha$-particle S1
pulses. The pure S1
contribution is reconstructed using an exponential curve fitted to the
initial part of the tail. The afterpulsing
contribution is obtained by subtracting the S1 contribution to the average
waveform.}
\end{figure}

As the PMT used in this work is old, we also took into account in its calibration the effect of the afterpulsing, generated by the ionization of residual gases inside the PMT and the subsequent drift of the positive ions towards the photocathode where they induce the emission of further electrons. The PMT afterpulsing contribution can be quantified using primary scintillation events, because they are sufficiently separated in time from the afterpulsing, given the short scintillation pulse of pure Xe, $\sim$100 ns decay time for the dimer triplet state. Figure~\ref{fig:afterpulsing}  depicts a typical waveform averaged over 1.5 $\times$ 10\textsuperscript{3} primary scintillation waveforms, obtained from $\alpha$-particle interactions. The reduced electric field in the absorption region was set to 140 V cm\textsuperscript{-1} bar\textsuperscript{-1} to prevent ion recombination \cite{10.1088/1748-0221/10/03/P03025, 10.1088/1748-0221/8/05/P05025, 10.1143/JJAP.48.076501, 10.1109/tns.2004.836030}, while keeping the neutral bremsstrahlung emission at residual levels \cite{NBrS_prx}. The primary scintillation waveforms were aligned using the rising edge at 50\% of the pulse’s height. The PMT afterpulsing is noticeable well separated from the primary scintillation tail.
An exponential function was fitted between 130 ns and 250 ns to avoid the contributions from both the fast xenon scintillation component of the singlet state and the afterpulsing. In this way, the afterpulsing signal could be obtained subtracting the reconstructed S1 pulse from the average waveform, as illustrated in Fig.~\ref{fig:afterpulsing}. An afterpulsing contribution of (24.9 \(\pm \ \)1.3) \% was measured. This value was crosschecked using waveforms obtained from a fast pulsed LED, attaining a 6\% lower value, relative to the former result.

We note that the results of Fig.~\ref{fig:afterpulsing} allow us to infer upper limits for the impurity content in our chamber \cite{10.1016/j.nima.2017.08.049, 10.1016/j.physletb.2017.09.017, 10.1007/JHEP01(2019)027}. From the above fit, a value of \(\tau_{3} =\) 108 \(\pm\) 5 ns was obtained for the decay time of the Xe dimer triplet state, to be compared with an average reference value of \(\tau_{3} =\) 100.9 \(\pm\) 0.7 ns \cite{10.1063/1.454313}. Using Eq. (1) from \cite{10.1016/j.nima.2017.08.049} together with the two-body quenching rates for excited xenon atoms reported in the literature, N\textsubscript{2}, CO\textsubscript{2}, O\textsubscript{2}, and CH\textsubscript{4} from \cite{10.1063/1.436447} and H\textsubscript{2}O from \cite{10.1021/j100211a041}, an upper limit of 21 ppm can be inferred for H\textsubscript{2}O, O\textsubscript{2}, CO\textsubscript{2}, and CH\textsubscript{4} concentrations at 95\% confidence level, and 409 ppm for N\textsubscript{2}. Lower values can be derived for heavier molecules based on the same references. Therefore, our absolute measurements of primary or secondary scintillation yields are unlikely to be biased by impurity quenching, since higher concentrations are typically required to have a sizable impact on Xe scintillation \cite{10.1016/j.nima.2017.08.049, 10.1016/j.physletb.2017.09.017, 10.1007/JHEP01(2019)027}.

\subsection{X-ray and $\gamma$-ray runs}
\label{sec:xrays}

\subsubsection{Energy and pulse duration cuts}
\label{sec:cuts}

\begin{figure}[tbp]
\centering % \begin{center}/\end{center} takes some additional vertical space
\includegraphics[width=12cm]{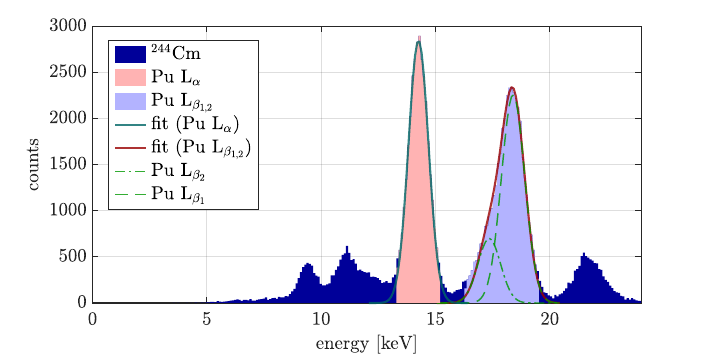}
\caption{\label{fig:Ecut} X-ray energy spectrum from a \textsuperscript{244}Cm
radioactive source as obtained from the S2 charge distribution. The
selected events of the 14.3- and 18.0-kev peaks (Pu \(L_{\alpha}\) and Pu
\(L_{\beta_{1,2}}\)) are depicted together with the respective Gaussian
fits. A double Gaussian function was used for the 18-keV peak. The 14.3-keV Gaussian fit was used for energy
calibration.}
\end{figure}

Secondary scintillation pulses were integrated, allowing to build the
energy spectrum of the radioactive sources, being the highest peak used as reference for energy calibration. The reconstructed
energy spectra have shown good linearity, with a deviation lower than 3\% from the different theoretical values. In this way,
waveforms originated for different energies can be selected for
subsequent S1 measurements. The peaks of interest were fitted to Gaussian functions, and events within
1.4 sigma, \(\sigma\), with respect to their centroids, \(c\), were
selected. A
double Gaussian fit was applied when the energy
peak is highly asymmetrical due to the presence of two different
energies, e.g. \(L_{\beta 1}\) and \(L_{\beta 2}\) lines, with both the ratio between the
centroids, \(c\), and the \(\sigma\) dependence on \(\sqrt{c}\) being
fixed. Accordingly, in those latter cases, the selected energy ranged from \(c_{1}\ –1.4\,\sigma_{1}\), of the first Gaussian,
to \(c_{2} + 1.4\,\sigma_{2}\), of the second Gaussian. The
ratio between the areas of the two fitted Gaussian functions was 
used to estimate the weighted theoretical energy of the selected double
peak, required for further calculations. Figure~\ref{fig:Ecut} 
shows two examples of energy cuts performed for the x-ray energy
spectrum of a \textsuperscript{244}Cm radioactive source, including the
Gaussian fits used for calibration and energy cuts.

\begin{figure}[tbp]
\centering % \begin{center}/\end{center} takes some additional vertical space
\includegraphics[width=12cm]{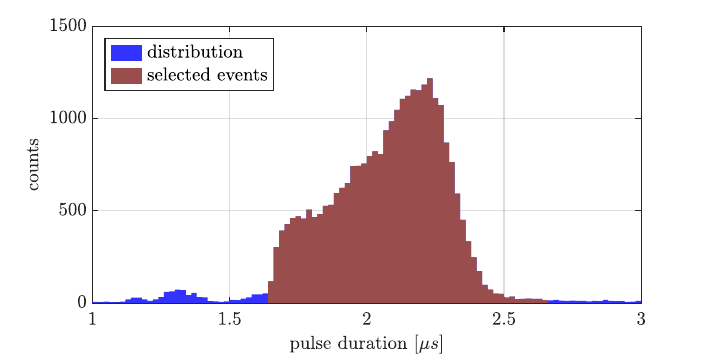}
\caption{\label{fig:Timecut} Distribution of the S2 pulse width of
selected 14.3-keV events. The S2 duration cuts are also illustrated.}
\end{figure}

Since S2 pulse
duration is directly proportional to the path length drifted by the primary
electron cloud along the EL region, this
parameter was used to further discriminate the waveforms.
Figure~\ref{fig:Timecut} 
shows the distribution of S2 pulse duration, measured
between the 50\%-thresholds of the S2 rising- and falling-edges, obtained from 14.3-keV interactions that passed in the energy
cut, Fig.~\ref{fig:Ecut}. 
The shape of the distribution is attributed to the longitudinal electron
diffusion: the longer the path length drifted by the electrons in the
absorption region, the larger the electron cloud, therefore
producing longer S2 pulses. The pulse width selection
region was defined to be between the 15\% and 1.5\% thresholds of the
rising and falling edges, respectively, as illustrated in Fig.
\ref{fig:Timecut}. The lower cut discriminates events corresponding to x-ray interactions
occurring inside the EL region, e.g. from x-ray interactions with higher energies, while the higher cut rejects some background
and anomalous waveforms.

\begin{figure}[tbp]
\centering % \begin{center}/\end{center} takes some additional vertical space
\includegraphics[width=9cm]{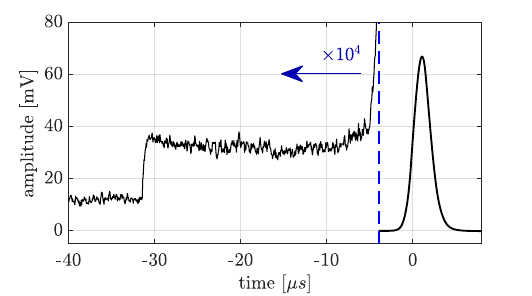}
\caption{\label{fig:averageW} Average waveform obtained for 14.3-keV x-rays
after energy and time cuts have been applied. The amplitude of the S1
continuum is zoomed-in by a factor of 10\textsuperscript{4}.}
\end{figure}

Finally, waveforms passing both the energy and the pulse duration cuts are
averaged to cancel out the electronic noise. The 50\%-threshold of the
S2 rising edge was chosen for the alignment of waveforms avoiding the
jitter introduced by S2 pulse-width fluctuations. Figure~\ref{fig:averageW} 
shows an average waveform obtained for 14.3-keV x-rays,
being computed from the events sampled in Fig. \ref{fig:Ecut} and Fig.
\ref{fig:Timecut}. Owing the large x/$\gamma$ -ray penetration along the absorption region, a S1 continuum is unveiled, due to interactions occurring at different depths, despite being 4 order of magnitude weaker when compared to S2. This continuum can’t be separated from the S2 rising tail.

\subsubsection{Spatial cuts}
\label{sec:vd}

\begin{figure}[tbp]
\centering\centering\centering
\includegraphics[width=9cm]{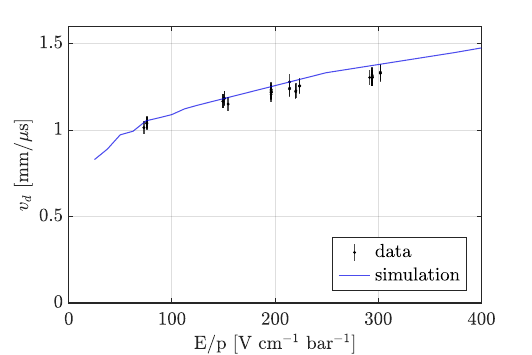}
\caption{\label{fig:driftSim} Electron drift
velocity values obtained experimentally compared with the \textsc{Magboltz}
simulation curve.}
\end{figure}

To avoid contaminating S1 measurements with the S2 contribution,
the primary scintillation yield obtained from the average waveforms was
determined selecting only x/$\gamma$-ray interactions occurring in the first
few centimetres of the absorption region. These spatial cuts were defined
using the electron drift velocity, which was measured for all the studied
energy peaks and all the reduced electric fields applied to the drift region. Electron drift velocities were computed from the time
elapsed between S1 and S2 rising edges of the average waveform,
corresponding to the transit time of electrons across the full
absorption region.  Figure~\ref{fig:driftSim} shows the electron drift velocity
obtained for all studied peak energies as a function of reduced
electric field. Our experimental data agree within two sigma with the
theoretical curve obtained from a \textsc{Magboltz} simulation, which is also depicted.

The electron drift velocity can be used to plot the waveform as a function of distance travelled by the primary electron cloud in the drift region. The integration of the S1 continuum of the average waveform was performed for the first 2.5 cm of the absorption region. This region was chosen to maximize the S1 statistics while keeping possible S2 contribution negligible.

\subsubsection{S1 yield determination}
\label{sec:xS1}

\begin{figure}[tbp]
\centering % \begin{center}/\end{center} takes some additional vertical space
\includegraphics[width=12cm]{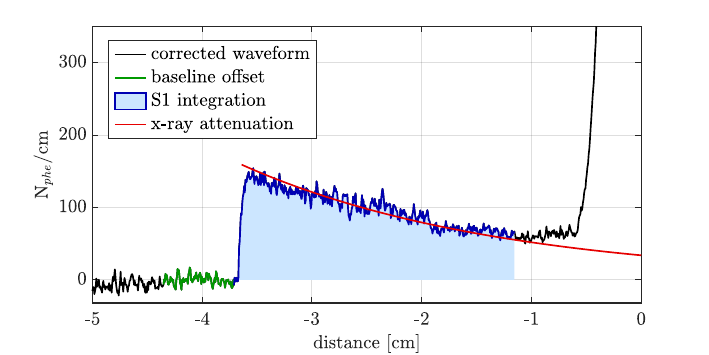}
\caption{\label{fig:corrW} Average waveform obtained for 14.3-keV x-rays
and corrected for the GE curve, plotted as a function of depth
where the primary scintillation occurs, with -3.6 cm and 0 cm
corresponding to the detector window and gate mesh, respectively. The
regions of interest used to compute the baseline offset, already
subtracted, and the S1 yield are depicted together with the theoretical
absorption curve of 14.3-keV x-rays in the 1.2-bar xenon gas.}
\end{figure}

The waveform of Fig.~\ref{fig:averageW} can be, now, corrected for the
detector geometrical efficiency, section \ref{sec:simulation}, 
to estimate the isotropic primary scintillation emission. Figure~\ref{fig:corrW} shows a
typical average waveform obtained for 14.3-keV x-rays, corrected for the
GE curve, in terms of the depth where the scintillation occurs: -3.6 cm
and 0 cm correspond to the detector window and gate mesh positions,
respectively. The baseline offset is measured immediately before the S1
integration region and subtracted to the waveform.

As shown, the GE-corrected waveform follows the theoretical exponential
x/$\gamma$-ray absorption law, which is also plotted in Fig.~\ref{fig:corrW}. 
This observation supports the reliability of the developed GE
simulation model. 

The average waveform was composed by all events under a specific energy
peak that occurred inside the whole absorption region. Therefore, the
ratio of events in the S1 integration region to the total
number of S1 events occurring in the absorption region had to be
accounted for in the determination of the primary scintillation yield. This parameter, \(R_{e}\),
was estimated from the exponential absorption law of x/$\gamma$-rays in xenon
expected for the respective experimental working conditions, i.e. the
theoretical energy of the impinging x/$\gamma$-ray, the gas pressure and the
temperature.

From the integration of the waveform amplitude, the total charge
induced in the PMT anode by the x/$\gamma$-ray interactions was determined. This charge was
converted into number of photoelectrons produced in the PMT photocathode
from the single-photoelectron mean charge, \(pe\), corrected for the
afterpulsing contribution, as obtained in section \ref{sec:PMTgain}.
Therefore, the number of primary scintillation photons per
event \(N_{ph}\) was calculated from the amplitude of the corrected average
waveform, \(U_{m}\), according to the following equation:
\begin{equation}
\label{eq:xNph}
N_{ph} = \frac{1}{R_{e}\ pe\ QE}\int_{-3.6cm}^{-1.1cm}\frac{U_{m}}{v_{d}\ GE}\ dz ,
\end{equation}
where \(QE\) is the PMT quantum efficiency at 175 nm, \(v_{d}\) is the
experimental electron drift velocity, $U_{m}/v_{d}$ the waveform amplitude and
\(GE\) the geometrical efficiency, both as a
function of distance to the detector window \(z\), integrated
over the -3.6 to -1.1 cm interval.

The energy required to produce one scintillation photon \(w_{sc}\) was obtained assuming the theoretical deposited energy \(E_{dep}\) of
the peak being studied or, in the case of
double peaks, the theoretical weighted energy:
\begin{equation}
\label{eq:wsc}
w_{sc} = \frac{E_{dep}}{N_{ph}}\ .
\end{equation}

\subsubsection{S2 yield determination}
\label{sec:xS2}

In addition to the primary scintillation determination, our setup allows
measuring the electroluminescence yield. This parameter was calculated using
the average charge of the S2 signal, \(c_{S2}\), obtained from the
centroid of a Gaussian fitted to the selected energy peak of the S2
charge distribution, as can be seen in Fig.~\ref{fig:Ecut}. In this way, the number
of EL photons produced per drift path length, per ionization electron
and per unit of pressure, \(Y_{el}\), was computed according to the equation:

\begin{equation}
\label{eq:Yel}
Y_{el} = \frac{c_{S2}}{\ pe\ QE\ \overline{GE_{el}}\ \Delta z\ P\ \overline{N_{e}}}\ ,
\end{equation}
where \(\overline{GE_{el}}\) is the average geometrical efficiency in
the EL region, 31\%, see Fig.~\ref{fig:GE}; \(\Delta z\) is
the EL gap length; \(P\) is the gas pressure, \(\sim\)1.2 bar;
and \(\overline{N_{e}}\) is the mean number of ionization electrons
produced in a x/$\gamma$-ray interaction. \(\overline{N_{e}}\  = E_{dep}/w_{i}\) where \(w_{i}\) is the
mean energy required to create one electron-ion pair in xenon. A
\(w_{i}\)-value of 21.6 \(\pm\) 0.2 eV was considered, as measured experimentally for xenon at a pressure around 1
bar in \cite{10.1109/TNS.2008.2003075}.

As a cross-check, the \(c_{S2}\) parameter can also be obtained from the
integration of the S2 pulse average waveform, as was done for the
primary scintillation analysis method. The \(Y_{EL}\) obtained for both the S2 charge distribution method and the average waveform integration method
were found to agree within a 3\%-error, supporting the reliability
of the average waveform analysis used in primary scintillation studies.
The former method for \(Y_{el}\) calculations has been henceforth applied, the difference between the two methods being included in the
experimental uncertainty.

\subsection{Alpha-particle runs}
\label{sec:alphaS1}

\begin{figure}[tbp]
\centering % \begin{center}/\end{center} takes some additional vertical space
\includegraphics[width=12cm]{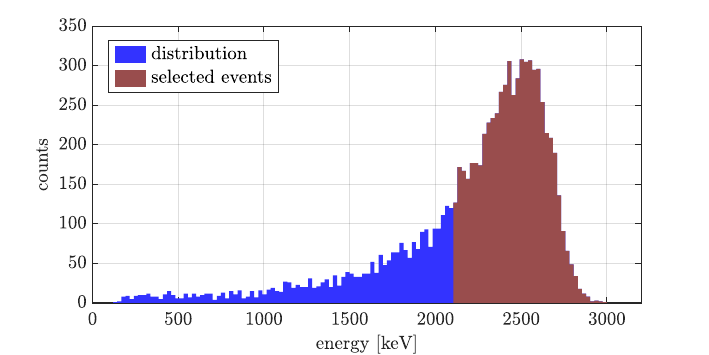}
\caption{\label{fig:alphaE} Energy spectrum for \textsuperscript{241}Am
$\alpha$-particles obtained from the S2 charge distribution. The energy cut is
also shown. Despite the fact that no Mylar foils have been used in this case to
further reduce $\alpha$-particle energies, a significant part of their energy was
lost to the detector window. The \textsuperscript{241}Am x/$\gamma$-ray peaks
are not visible due to the chosen oscilloscope trigger threshold, which was set
above the amplitude of these pulses.}
\end{figure}

The energy of \textsuperscript{241}Am $\alpha$-particles is 2 orders of
magnitude higher than the studied x/$\gamma$-ray energy range. To avoid PMT
saturation for S2 pulses, it was biased at a lower voltage, 800 V instead
of 1450 V. The PMT gain scaling factor was measured using the centroid
of the 5.9-keV x-ray peak from the \textsuperscript{55}Fe energy spectrum, acquired for
both PMT voltages and using the same reduced electric field values. The
theoretical energy deposited by $\alpha$-particles in GXe cannot be accurately
simulated, due to the unknown thickness of both the aluminium-deposited
film on the detector window and the gold protective layer of the
radioactive source. Therefore, the detector was calibrated with the
5.9-keV peak from the \textsuperscript{55}Fe radioactive source, using the same reduced
electric field values and gas pressure as in $\alpha$-particle runs. Figure~\ref{fig:alphaE}
shows a typical energy spectrum obtained for the S2 charge
distribution of $\alpha$-particles after calibration. The different layers of
materials degraded energy and trajectory of $\alpha$-particles before they
reached the xenon gas, resulting in a left-tailed peak. Low energy events were discriminated as they might have
occurred too close to the detector window where the electric field is
weaker, and some of the produced ionization electrons could have been lost to the
window electrode. Thus, only events with energies higher than 40\% of the height
of the energy peak were accounted for in primary scintillation calculations,
as illustrated in Fig.~\ref{fig:alphaE}. Pulse duration cuts were not
required for $\alpha$-particles because they did not reach the EL region.

\begin{figure}[tbp]
\centering % \begin{center}/\end{center} takes some additional vertical space
\includegraphics[width=12cm]{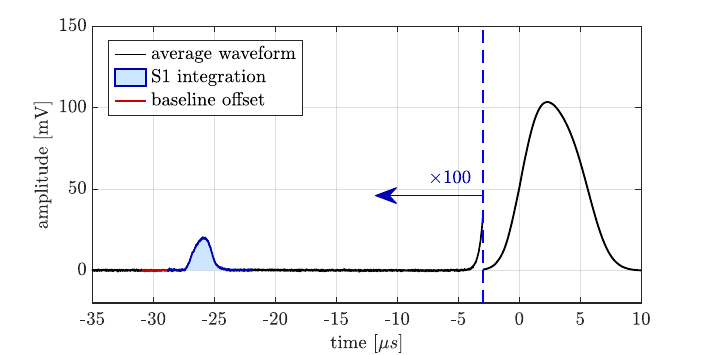}
\caption{\label{fig:alphaAve} Average waveform of selected 2.5-MeV
$\alpha$-particle events. The regions of interest used for baseline offset,
already subtracted, and the calculated S1 yield values are also shown.}
\end{figure}

As in the x/$\gamma$-ray analysis method, the average waveform was computed
from selected events and corrected for the baseline offset that is
measured before S1. Figure~\ref{fig:alphaAve} depicts a
typical waveform averaged over 6 $\times$ 10\textsuperscript{3} events.
Since the $\alpha$-particle
penetration is short and the electron cloud drift time is similar for all the
events, the S1 pulse is sharp and localized. Therefore, the primary
scintillation yield can be obtained directly from the integration over the
full S1 pulse, hence avoiding the need for spatial cuts. Figure~\ref{fig:alphaAve} illustrates as well the waveform regions used for
baseline offset correction and for S1 integration.

The GE correction was calculated from the simulated GE curve, Fig.
\ref{fig:GE}, weighted over the energy deposition along the $\alpha$-particle interaction depth, Fig.~\ref{fig:tracking} left, as obtained from \textsc{GEANT4} $\alpha$-particle
tracking, described in section \ref{sec:simS1}. Mylar films with different thicknesses were used
to degrade the $\alpha$-particle energies in order to irradiate the detector with
$\alpha$-particles of different energies. Accordingly, for every data
acquisition, the energy of simulated $\alpha$-particles was adjusted for the deposited energy distribution to match the mean energy measured
experimentally after the energy cuts were applied. The mean geometrical efficiency
value, \(\overline{GE}\), was found to be in the 6\% to 7\% range, depending on the $\alpha$-particle energy being studied. In
this way, the number of primary scintillation photons generated per
event, \(N_{ph}\), was computed as:
\begin{equation}
\label{eq:aNph}
N_{ph} = \frac{1}{\ pe\ QE\ \overline{GE}\ }\int_{- 2\ \mu s}^{5\ \mu s}U_{m}dt\ ,
\end{equation}
where \(U_{m}\) is the amplitude of the average waveform
integrated between -2 \(\mu s\) and 5 \(\mu s\) with respect to 50\% of the S1 pulse
rising edge. \(QE\) and \(pe\) are the same values as
in the x/$\gamma$-ray analysis method. Finally, the \(w_{sc}\)-value was
obtained from:
\begin{equation}
\label{eq:wsc}
w_{sc} = \frac{E_{dep}}{N_{ph}}\ .
\end{equation}
Where \(E_{dep}\) is the measured deposited
energy, averaged over the selected events, Fig.
\ref{fig:alphaE}.

\begin{figure}[tbp]
\centering\centering\centering
\includegraphics[width=7.5cm]{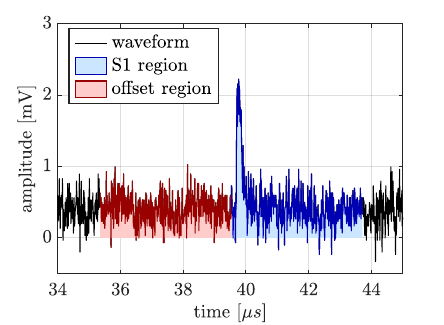}
\includegraphics[width=7.5cm]{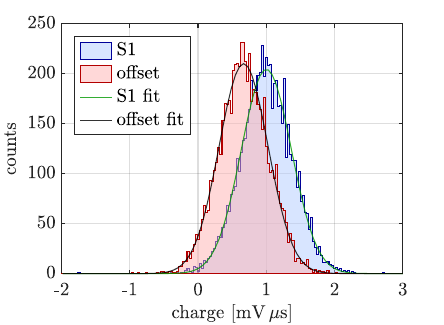}
\caption{\label{fig:alphaEvent} (Left) Waveform of a typical S1 pulse produced by
an $\alpha$-particle interaction. The integration regions considered for
baseline offset and S1 yield calculations are also depicted; (right)
charge distribution from the baseline and S1 regions of interest
obtained for 2.5-MeV $\alpha$-particles, along with the corresponding Gaussian
fits.}
\end{figure}

In contrast to x/$\gamma$-ray runs, the S1 pulse was large enough to be
detected in the individual $\alpha$-particle waveforms, allowing to perform
per-event statistics. For this analysis method the S1 pulse-time was
automatically detected as the maximum amplitude in the S1 region of the
waveform, which was previously processed with a software-implemented
differentiator and a moving average to remove both low- and high-frequency
fluctuations. Finally, the baseline offset and the S1 pulse were
integrated in the intervals -4.2 \(\mu s\ \)to -0.2 \(\mu s\) and -0.2
\(\mu s\) to 3.8 \(\mu s\), respectively, with respect to the S1 peaking
time. Figure~\ref{fig:alphaEvent} left shows an example of a single
$\alpha$-particle waveform with highlighted integration regions. Figure
\ref{fig:alphaEvent} right shows the distribution of integration values for
both baseline offset and S1 pulse. Despite the energy of the selected 
$\alpha$-particle events not obeying a Gaussian distribution, the S1 integral
distribution is roughly Gaussian due to the large number of statistical
fluctuations. Therefore, both S1 mean charge value and mean baseline
offset could be estimated from the centroid of a Gaussian function fitted
to each distribution, as shown in Fig.~\ref{fig:alphaEvent} right.

The difference between \(w_{sc}\)-values obtained from both methods,
average waveform and per-event statistics, was under 2\%. This agreement
supports our assumption that the waveform average method does not introduce
a meaningful systematic error in the results. 
\(w_{sc}\) results for $\alpha$-particles reported henceforth correspond to the mean value obtained with both methods.

\section{Results and discussion}
\label{sec:results}

\begin{figure}[tbp]
\centering % \begin{center}/\end{center} takes some additional vertical space
\includegraphics[width=12cm]{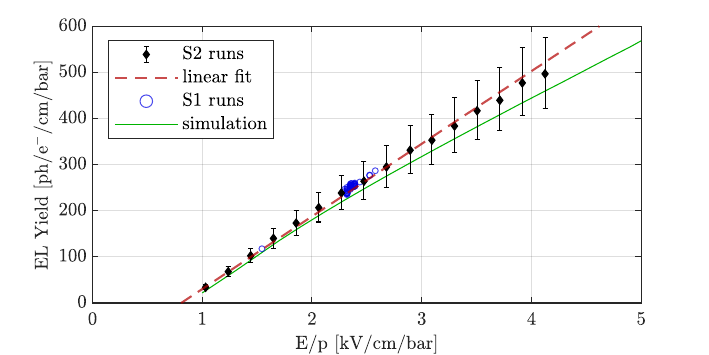}
\caption{\label{fig:EL} Experimental electroluminescence yield as a function of reduced electric field together with the corresponding linear fit. Yield values
obtained from S1 data runs are also shown, along with the \textsc{Garfield++}
simulation data.}
\end{figure}

The reduced electroluminescence yield, \(Y_{el}/p\), as a
function of reduced electric field, \(E/p\), in the
EL region is shown in Fig.~\ref{fig:EL}, together with the
theoretical curve obtained with \textsc{Garfield++} simulation,
as described in section \ref{sec:simS2}. 5.9-keV x-rays 
from a \textsuperscript{55}Fe radioactive source were used, although with much
lower acquisition times when compared with primary scintillation runs.
Only the systematic error is depicted since the statistical
uncertainty was lower than 3\%, thus negligible. 
EL yield values that were
obtained from the primary scintillation acquisition runs are depicted as well.
The typical approximately linear dependency of EL yield with electric field
is observed. A line fitted to the experimental data is also depicted,
having the following fit parameters:
\(Y_{el}/p = (157 \pm 4)\ E/p \ – \ (127 \pm 5)\), where \(Y_{el}/p\) and
\(E/p\) have units of \(ph/(e^{-}\ cm\ bar)\) and
\(kV/(cm\ bar)\), respectively. The interception of the fitted
line with the \(E/p\) axis, defined as the EL threshold is
$0.81 \pm 0.04$ \(kV/(cm\ bar)\). The experimental EL yield is about 7\%
higher than theoretical predictions and 14\% higher when compared to the
experimental values obtained in a driftless GPSC
\cite{10.1088/1748-0221/2/05/P05001}. Nevertheless, these differences are
within experimental uncertainties.

\subsection{\boldmath{$w_{sc}$} absolute measurements}
\label{sec:w_absolute}

The \(w_{sc}\)-value was measured for different electric field values in the
absorption and in the EL region to evaluate possible systematic errors and the
role of electron-ion recombination. In the absorption region, the
electric field was varied in the 70--300 V cm\textsuperscript{-1}
bar\textsuperscript{-1} range. At moderately high electric field values, the
recently unveiled neutral bremsstrahlung emission in xenon becomes
significant when compared with the weak primary scintillation emission.
We resorted to the NBrS photoelectron yield measured in
\cite{NBrS_prx} to estimate its contribution in
our data. The NBrS contribution in the S1 integration region was
estimated considering the geometrical efficiency model and the
distribution of event position along the absorption region. For the highest
electric field value, 300 V cm\textsuperscript{-1} bar\textsuperscript{-1},
the NBrS accounted for 11\% of the total S1 charge,
being negligible below 150 V cm\textsuperscript{-1}
bar\textsuperscript{-1}. These corrections were included in the
\(w_{sc}\) calculations. For $\alpha$-particle runs the NBrS
impact was negligible because only a short region of the waveforms was
integrated. We did not find any significant dependency of the \(w_{sc}\)-value, duly corrected for NBrS contribution, on the electric fields either in the absorption region or in the EL region, within the studied ranges. Therefore, any major systematic error associated to electric field is unlikely to occur.

\begin{figure*}[tbp]
\centering % \begin{center}/\end{center} takes some additional vertical space
\includegraphics[width=15cm]{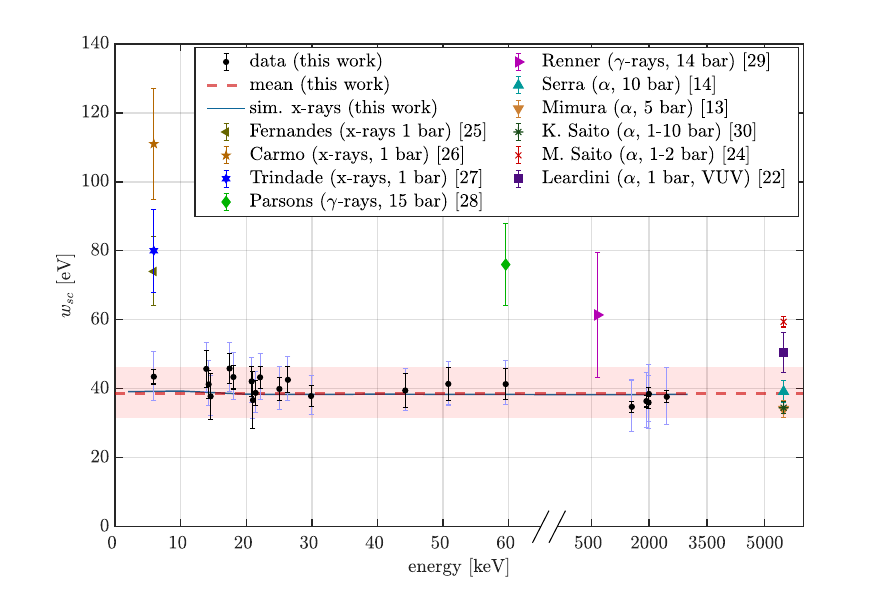}
\caption{\label{fig:wsc} Absolute \(w_{sc}\)-value obtained experimentally for
x-rays, $\gamma$-rays and $\alpha$-particles as a function of deposited energy; simulation and literature experimental data are included for comparison. The
statistical and systematic uncertainties in our experimental data are
presented in black and blue error bars, respectively, both referring
to the 68\% confidence level. The statistically combined
\(w_{sc}\)-value obtained from our experimental data is also shown,
the overall uncertainty being depicted as a red band. For all the data in the literature, the 3\textsuperscript{rd} continuum emission is assumed to be negligible, except for Leardini’s value.}
\end{figure*}

Figure~\ref{fig:wsc} shows the \(w_{sc}\)-values measured in the absence of
recombination for x/$\gamma$-ray and $\alpha$-particle interactions in the energy
range between 5.9 keV and 2.5 MeV, assuming a negligible
3\textsuperscript{rd} continuum emission. Data points corresponding to
the same energy were statistically combined for clarity. Systematic and
statistical errors at the 68\% confidence level are displayed with
separated error bars. The systematic uncertainty arises mainly from the
detector geometrical efficiency simulation, Fig.~\ref{fig:GE}, though
with sizeable contributions from the PMT single-photoelectron calibration
and quantum efficiency, 9\% and 7\%, respectively. Whereas the latter
two contributions are energy-independent, the GE uncertainty is generally
lower for high energy x/$\gamma$-rays, since the relevance of photon reflection
becomes smaller for interactions occurring closer to the PMT. This
partially explains the large systematic error in $\alpha$-particle runs, which also includes
an additional error source stemming from the detector energy calibration, 14\%.
The statistical uncertainty was
dominated by the oscilloscope's baseline fluctuations. Therefore, this error
was lower in $\alpha$-particle data due to the much higher S1 amplitude when compared
to the baseline fluctuations. 

The mean \(w_{sc}\)-value of our experimental data, 38.7 \(\pm\) 0.6
(sta.)\(\ _{- 7.2}^{+ 7.7}\) (sys.) eV, is also depicted in Fig.~\ref{fig:wsc},
being computed from the entire dataset weighted over the respective
statistical errors. When the dataset was split into x/$\gamma$-ray and $\alpha$-particle
runs, the mean \(w_{sc}\)-values obtained were 41.8 \(\pm\) 1.0
(sta.)\(\ _{- 6.2}^{+ 6.8}\) (sys.) eV and 36.6 \(\pm\) 0.8
(sta.)\(\ _{- 7.7}^{+ 8.2}\) (sys.) eV, respectively. These differences are within the
experimental uncertainty, preventing us from precisely observe any
fundamental distinction between the primary scintillation yield for
electrons and alpha particles.

All the values depicted in Fig. \ref{fig:wsc} assume the 3\textsuperscript{rd} continuum emission to be negligible, except the value reported
by Leardini \emph{et al.}, which refers to the
2\textsuperscript{nd} continuum only, as they used optical filters to suppress the
most part of the 3\textsuperscript{rd} continuum emission 
\cite{10.1140/epjc/s10052-022-10385-y}. However, the measured
values for the primary scintillation yield includes the contribution of
both the 2\textsuperscript{nd} and the 3\textsuperscript{rd} continuum, as the PMT is sensitive to wavelengths in the 150-650 nm region. Our PMT QE, Fig.~\ref{fig:QE}, was a factor of 1.32 higher for the
xenon 3\textsuperscript{rd} continuum with respect to the
2\textsuperscript{nd} continuum, while the GE of the detector was a factor
of 1.05 higher, as obtained from the GE simulation model. At 1.2 bar,
the ratio between the contributions from the 3\textsuperscript{rd} and the 2\textsuperscript{nd}
continuum to the primary scintillation emission in the
absence of recombination was measured to be 0.09 \(\pm\) 0.01
\cite{10.1140/epjc/s10052-022-10385-y}. Therefore, the mean energy
required to produce a 2\textsuperscript{nd} continuum scintillation
photon \(w_{2^{nd}}\) was estimated to be 
\(w_{2^{nd}} = (1\ +\ 1.32 \times 1.05 \times 0.09)\ w_{sc}\) = 43.5 \(\pm\) 0.7
(sta.)\(\ _{- 8.1}^{+ 8.7}\) (sys.) eV, where
\(w_{sc}\) is the measured value when neglecting the
3\textsuperscript{rd} continuum emission. 
The mean energy required to
produce a 3\textsuperscript{rd} continuum photon was computed as
\(w_{3^{rd}} = ( 1/0.09 )\ w_{2^{nd}} =\) 483
\(\pm\) 7 (sta.)\(\ _{- 105}^{+ 110}\) (sys.) eV and the mean
energy required to produce a primary scintillation photon was obtained as
\(w_{2^{nd}{+ 3}^{rd}} = \ w_{2^{nd}}/(1 + 0.09) =\) 39.9
\(\pm\) 0.6 (sta.)\(\ _{- 7.4}^{+ 8.0}\) (sys.) eV.

As a crosscheck, instead of relying on the PMT single photoelectron response and the absolute simulated GE for optical calibration of the detector, one could use the S2 charge distribution and the corresponding yield value given by simulation. Since the experimental EL yield depicted in Fig.~\ref{fig:EL} is 7\% higher than the simulation value, one may argue that data obtained by the former method are overestimated, e.g. due to a systematic uncertainty related to the PMT calibration. Accordingly, the \(w_{sc}\)-values obtained with that method could be underestimated. Assuming this second method, the results obtained with the former method could be normalized considering the ratio between theoretical and experimental EL yields, as the latter was measured for every run by integrating the S2 pulses of the waveforms. Such a correction would make \(w_{sc}\)-values 7\% higher on average, which nevertheless is still within the 68\% confidence level of the present data of Fig. \ref{fig:wsc}. If, on one hand, this second analysis method has the advantage of eliminating the systematic uncertainty introduced by both PMT calibration and QE, as well as a sizeable part of the GE uncertainty. On the other hand, it introduces new error sources, such as the width of the EL region and the gas temperature, both contributing \(\sim\)5\%. For this reasons, we consider the absolute \(w_{sc}\)-values depicted in Fig. \ref{fig:wsc} to be more accurate than EL-corrected values.

As another crosscheck to our \(w_{sc}\)-values we applied the same analysis method used in this work to other two gas scintillation chambers filled with Xe, namely a driftless GPSC used in
\cite{NBrS_prx, 10.1016/j.physletb.2017.09.017, 10.1007/JHEP01(2019)027} and a similar chamber built for Kr-gas studies
\cite{daniel}. The \(w_{sc}\)-values obtained for
5.9-keV x-rays and 2-MeV $\alpha$-particles agree within experimental uncertainties with the values presented here. 

\subsection{\boldmath{$w_{sc}$} comparison with literature values}
\label{sec:literature}

Figure \ref{fig:wsc} illustrates well how dispersed is
the data from the literature. Despite the theoretical values, around 40 eV, are
compatible with most of the $\alpha$-particle results, 
there is a clear disagreement with x/$\gamma$-ray data. \(w_{sc}\) is expected to be similar for x-, $\gamma$-rays or electrons and almost equal to that obtained for $\alpha$-particles \cite{10.1143/JJAP.48.076501}. However, the results presented so far in the literature are inconsistent with that expectation, being this difference not understood.
The \(w_{sc}\)  results presented in this work agree with both theoretical
predictions and most of $\alpha$-particle experimental values. In addition, the present results do not show a dependency of \(w_{sc}\) with the nature or the energy of the impinging radiation, being incompatible with literature's values obtained for x/$\gamma$-rays, even considering the experimental uncertainties.
The different experimental conditions reported in the literature seem insufficient to explain
the discrepancies. The impact of gas pressure is expected to be
minor up to 20 bar \cite{10.1016/j.nima.2017.08.049,
10.1143/JJAP.48.076501, 10.1140/epjc/s10052-022-10385-y} and for
electric field values above 60 V cm\textsuperscript{-1}
bar\textsuperscript{-1}, recombination is negligible \cite{10.1088/1748-0221/10/03/P03025, 10.1088/1748-0221/8/05/P05025, 10.1143/JJAP.48.076501, 10.1109/tns.2004.836030}.

In order to clarify this puzzle, we attempted to replicate
the experimental methodology of Fernandes' and Carmo's works 
\cite{10.1088/1748-0221/3/07/P07004, 10.1088/1748-0221/5/09/P09006}, since
their experimental apparatus were similar to the one we have used. A major difference was the fact that the data acquisition
system in those works consisted of conventional x-ray spectroscopy electronics, i.e.
instead of feeding the PMT signal directly into the oscilloscope, the
PMT signal was first integrated and shaped by a pre-amplifier and a
linear amplifier chain. Like in our analysis method, the weak primary
scintillation signal was unveiled by averaging several PMT waveforms,
though with far less statistics, 128 against 10\textsuperscript{6} events. 
Due to hardware signal integration, the S1 and S2 heights were
proportional to the PMT charge produced by the primary and secondary
scintillation emissions. Hence, the S1 and S2 pulse heights from 5.9 keV x-rays were measured
from two averaged waveforms acquired in different runs: one with a low
amplification level set in the linear amplifier to avoid S2
pulse saturation, and the other with high amplification to reveal the S1
pulses. The ratio between the two pulse heights was corrected for the
difference between the solid angles subtended by the PMT with respect to
the primary and secondary scintillation emissions, being the absolute
EL yield established in the literature used to estimate the total
number of primary scintillation photons, in spite of the 4 orders of magnitude difference between them.

We were able to reproduce Fernandes' and Carmo's findings by replicating
the above methodology. We identified several issues in the analysis method leading to large
systematic errors, which in our opinion were not properly accounted for in those
works. The lack of an adequate light propagation model to quantify the
geometrical efficiency in the absorption and in the EL region was the most obvious one.
Figure~\ref{fig:GE} illustrates how important photon reflection is. From our simulation
data, a $\sim$50\% underestimation of the \(w_{sc}\)-value would be expected,
yet this is the opposite that is shown in Fig. \ref{fig:wsc}. 
Therefore, a much larger and opposite experimental error would be required to explain those results.

A major source of systematic errors is the oscilloscope trigger
threshold, which serves in this analysis as the only method of event
selection and discrimination.
Since S1 and S2 pulses were averaged over every waveform, their heights
include all background events occurring above the trigger threshold,
such as cosmic rays. Indeed, we observed a $\sim$100\% increase of the
\(w_{sc}\) parameter just by tightly collimating the \textsuperscript{55}Fe radioactive
source, thus increasing the ratio between background and 5.9-keV x-ray
events. This finding is disruptive, since it demonstrates how the
\(w_{sc}\) measurement can be easily biased by the radioactive source activity
and by the background levels of a given experiment. When measuring the S1 height of the
average waveform, the signal needs to be amplified by about 3 orders of
magnitude by the linear amplifier. Consequently, the maximum setting of the oscilloscope's trigger
threshold is much lower than S2
amplitudes. In such conditions, all S2 pulses with energies
above \(\sim\)0.0059 keV are triggered, and even some S1 pulses can be
triggered as if they were S2 pulses. Many of these backgrounds or badly
sampled waveforms do not actually have any photon emission in the S1
region, thus leading to an underestimation of the average S1 height.
Indeed, we found out that by simply decreasing the trigger threshold, the
\(w_{sc}\)-value could increase from 60 eV to 150 eV.

\section{Conclusions}
\label{sec:conclusion}

We have carried out an experimental campaign to measure the
gaseous xenon primary scintillation yield, \(w_{sc}\) -- the average
energy to produce a primary scintillation photon, for interactions of
alpha particles in the 1.5 -- 2.5 MeV range and for x/$\gamma$-rays in the 6 to
60 keV range. We used a gas scintillation chamber
instrumented with a PMT to readout both the primary scintillation, S1,
produced upon radiation interaction and the secondary proportional
scintillation, S2, produced by the primary ionization electrons. For
the purpose we digitized the PMT waveforms using the large S2 pulses to
trigger the digitizer, and averaged a very large number of pulses,
\(\sim\)10\textsuperscript{6}, to cancel out the electronic noise and unveil the signal from the primary scintillation.

Two methods have been used: optical calibration through single photoelectron response and
optical calibration using the area of the S2 waveform being the corresponding yield value given by simulation. Both methods agree within the experimental uncertainties.

Measuring the \(w_{sc}\)-value is far more challenging for low energy
x-rays than for $\alpha$-particles. Whereas per-event statistics can be used in
$\alpha$-particle runs, waveforms averaging is required for x/$\gamma$-rays. We validated the latter method by comparing
it with per-event statistics for $\alpha$-particles, achieving similar results well within the experimental uncertainties. 

We obtained a \(w_{sc}\)-value that is simultaneously compatible with
well-established literature data reported for $\alpha$-particles and in good
agreement with state-of-the-art simulations performed for x/$\gamma$-ray
interactions. In addition, the present results do not show a significant dependency of \(w_{sc}\) with the nature or the energy of the impinging radiation. These arguments sustain our hypothesis that some literature
\(w_{sc}\)-values obtained for x/$\gamma$-ray are unreliable due to undressed
systematic errors.

The results obtained in this work are summarized in Table \ref{tab:data} for the absolute
\(w_{sc}\)-value. We chose to present the mean
\(w_{sc}\)-value of all our experimental data, being computed from the
entire dataset weighted for the respective statistical errors. In addition, we present the
\(w_{sc}\)-value for second and third continua separately,
assuming a 3\textsuperscript{rd}-to-2\textsuperscript{nd} continuum yield ratio of 0.09, as recently disclosed in the literature, and the \(w_{sc}\)-value for the emission of either 3\textsuperscript{rd} or 2\textsuperscript{nd} continuum photon.

\begin{table*}[tbp]
\setlength{\tabcolsep}{11pt}
\renewcommand{\arraystretch}{1.3}
\centering
\begin{tabular}{c c c c}
\hline \hline
\bf \boldmath $w_{sc} $ (eV) &  \bf observations \\
\hline
38.7 \(\pm\) 0.6 (sta.)\(\ _{- 7.2}^{+ 7.7}\) (sys.) & 3\textsuperscript{rd} continuum neglected\\
43.5 \(\pm\) 0.7 (sta.)\(\ _{- 8.1}^{+ 8.7}\) (sys.) & 2\textsuperscript{nd} continuum \\
483 \(\pm\) 7 (sta.)\(\ _{- 105}^{+ 110}\) (sys.) & 3\textsuperscript{rd} continuum \\
39.9 \(\pm\) 0.6 (sta.)\(\ _{- 7.4}^{+ 8.0}\) (sys.) & 2\textsuperscript{nd}+3\textsuperscript{rd} continua \\
\hline
\hline
\end{tabular}

\caption{\label{tab:data}A summary of the mean \(w_{sc}\)-values measured in the present work for x-rays, $\gamma$-rays and $\alpha$-particles in the 6--2500 keV energy range, either considering or neglecting the Xe 3\textsuperscript{rd} continuum. }
\end{table*}

\acknowledgments
This work was fully funded by national funds, through FCT -- Fundação para a Ciência e a Tecnologia, I.P., under Projects No. UIDP/04559/2020, UIDB/04559/2020, CERN/FIS-TEC/0038/2021 and PTDC/FIS-NUC/3933/2021, and under Grant No. UI/BD/151005/2021.
We thank C.D.R Azevedo for introducing to simulation inter-using \textsc{Magboltz}, \textsc{Garfield++} and \textsc{Degrad}. We thank D. Gonzalez-Dias for fruitful discussions.

% Bibliography

%% [A] Recommended: using JHEP.bst file
%% \bibliographystyle{JHEP}
%% \bibliography{biblio.bib}

%% or
%% [B] Manual formatting (see below)
%% (i) We suggest to always provide author, title and journal data or doi:
%% in short all the informations that clearly identify a document.
%% (ii) please avoid comments such as "For a review'', "For some examples",
%% "and references therein" or move them in the text. In general, please leave only references in the bibliography and move all
%% accessory text in footnotes.
%% (iii) Also, please have only one work for each \bibitem.

\bibliographystyle{JHEP}
\bibliography{biblio.bib}

\end{document}